\documentclass[acmsmall]{acmart}

\usepackage{graphicx}
\usepackage{textcomp}
\usepackage{xcolor}
\usepackage{multirow}
\usepackage{xspace}
\usepackage{tcolorbox}
\usepackage{amsmath}
\usepackage{listings}
\usepackage{algorithm}
\usepackage{subcaption}
\usepackage{algpseudocode}
\usepackage{float}       
\usepackage{booktabs}
\usepackage{wrapfig} 
\usepackage{graphicx}
\usepackage{soul}
\tcbuselibrary{listingsutf8}
\definecolor{codegray}{rgb}{0.97,0.97,1}
\definecolor{commentgreen}{rgb}{0.1,0.5,0.1}
\definecolor{keywordblue}{rgb}{0.2,0.2,0.7}
\definecolor{stringred}{rgb}{0.8,0.1,0.1}

\newtcblisting{problemcodebox}{
  listing only,
  colframe=black!60,
  fonttitle=\bfseries,
  title=An Example from MathQA,
  label={lst:mathqa},
  listing options={
    basicstyle=\ttfamily\footnotesize,
    showstringspaces=false,
    breaklines=true,
    tabsize=2
  },
}
\AtBeginDocument{%
  }

\newcommand{\techname}{GramTrans\xspace}
\begin{document}

\title{GramTrans: A Better Code Representation Approach in Code Generation}
\author{Zhao Zhang}
\affiliation{%
  \institution{Peking University}
  \city{Beijing}
  \country{China}
}
\email{zhangzhao2019@pku.edu.cn}

\author{Qingyuan Liang}
\affiliation{%
  \institution{Peking University}
  \city{Beijing}
  \country{China}
}
\email{liangqy@pku.edu.cn}

\author{Zeyu Sun}
\affiliation{%
  \institution{Institute of Software, Chinese Academy of Sciences}
  \city{Beijing}
  \country{China}
}
\email{zeyu.zys@gmail.com}

\author{Yizhou Chen}
\affiliation{%
  \institution{Peking University}
  \city{Beijing}
  \country{China}
}
\email{yizhouchen@stu.pku.edu.cn}

\author{Guoqing Wang}
\affiliation{%
  \institution{Peking University}
  \city{Beijing}
  \country{China}
}
\email{guoqingwang@stu.pku.edu.cn}

\author{Yican Sun}
\affiliation{%
  \institution{Peking University}
  \city{Beijing}
  \country{China}
}
\email{sycpku@pku.edu.cn}

\author{Lu Zhang}
\affiliation{%
  \institution{Peking University}
  \city{Beijing}
  \country{China}
}
\email{zhanglucs@pku.edu.cn}

\author{Ge Li}
\affiliation{%
  \institution{Peking University}
  \city{Beijing}
  \country{China}
}
\email{lige@pku.edu.cn}

\author{Yingfei Xiong}
\affiliation{%
  \institution{Peking University}
  \city{Beijing}
  \country{China}
}
\email{xiongyf@pku.edu.cn}








\renewcommand{\shortauthors}{Zhang et al.}

\begin{abstract}


Code generation has shown great promise in assisting software development. A fundamental yet underexplored question is how the choice of code representation affects model performance. While existing studies employ various representations, such as treating code as plain text, grammar rule sequences, or syntax tree sequences, they lack a principled understanding of the relationship between parsing difficulty and model effectiveness.

This paper proposes a conjecture: the easier a representation is to parse, the better performance the model achieves. We formalize this idea using grammar classes, where representations in simpler classes (e.g., LL(1)) are easier to parse. Through a controlled experiment on a Python-based DSL, we show that parsing difficulty strongly correlates with model performance. Motivated by this finding, we present \techname, a general approach that automatically transforms a context-free language into a representation within the LL(1) class. \techname introduces a novel hierarchical conflict elimination algorithm, enabling a flexible trade-off between syntactic simplicity and token efficiency.

We evaluate \techname on both Python and Java using three code generation models: StarCoder~1B, DeepSeek-Coder~1.3B, and Qwen2.5~1.5B. Across multiple benchmarks, \techname consistently delivers significant improvements over baseline representations. Furthermore, our analysis of existing representations reconfirms the strong alignment between parsing difficulty and model performance, providing additional support for the conjecture.

\end{abstract}
\settopmatter{printacmref=false}
\renewcommand\footnotetextcopyrightpermission[1]{}
\makeatletter
\def\@copyrightpermission{}
\makeatother
\maketitle
\section{Introduction}

Code generation has emerged as a promising approach to improve software productivity and has attracted significant research attention in recent years~\cite{zhu2024deepseek,balog2016deepcoder,feng2020codebert,chatgpt3.5}. 
A key design consideration is how to represent programs during model training and inference. The choice of representation can influence the model’s ability to capture syntactic patterns and semantic relationships, ultimately affecting the effectiveness and efficiency of code generation.

While some work represents programs directly as plain text~\cite{bpe}, consistent with natural language, many studies have explored alternative representations, most of which try to utilize structural information.
These approaches can be typically divided into three groups: (i)~grammar-rule-based approaches represent programs as sequences of grammar rules, thereby reflecting the construction of the syntax tree~\cite{sun2019grammar,sun2020treegen,grammart5,liang2025grammar}; (ii)~syntax-tree-based approaches represent programs as traversals of syntax trees, exposing internal nodes to capture structural information~\cite{jiang2021treebert,guo2022unixcoder,wang2021syncobert,sbt}; (iii)~specially designed programming language approaches translate programs into new languages, removing redundant information from the code structure to shorten the representation length~\cite{simpy}. 

\begin{figure}
\centering
        \includegraphics[width=0.85\textwidth]{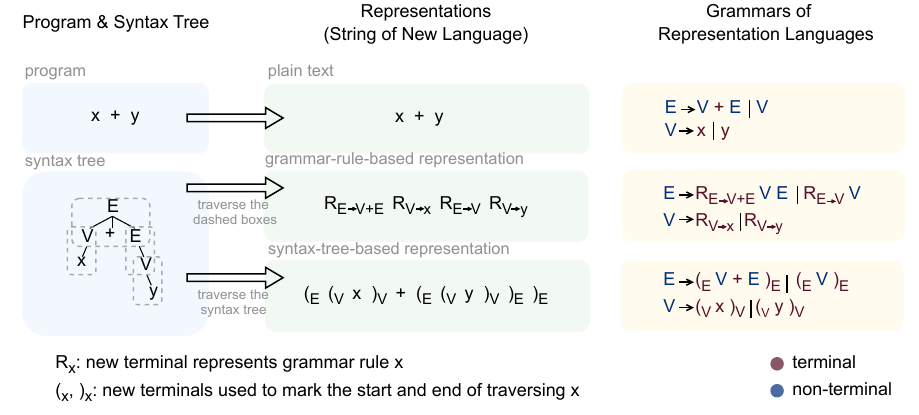}
        \caption{Different representations of the program example. The program and its syntax tree are shown on the left, and different representations are illustrated in the middle. The grammar-rule-based representation is derived from the traversal of the dashed boxes~(grammar rules), and the syntax-tree-based representation is obtained by traversing the syntax tree. These representations can themselves be viewed as new languages, whose grammars are given on the right.}
        \label{figure:intro_example}
\end{figure}

It is worth noting that, despite the diversity of these representations, the final inputs to models are still strings. Figure~\ref{figure:intro_example} illustrates an example: the left column shows the program \texttt{x+y} along with its syntax tree, and the middle column presents two designed representations~(i.e., the grammar-rule–based representation~\cite{grammart5}, and the syntax-tree–based representation~\cite{sbt}) beyond the plain text representation. Though these representations extract syntax trees from the code, the syntax trees are ultimately reduced to strings as model input and output. This practice rests on a basic assumption: the model possesses an implicit ability to parse such strings and recover the underlying structural information. Building upon this assumption, we argue that the parsing difficulty of strings can significantly impact model performance. In other words, while structural information is important, its effectiveness may depend not only on whether it is provided but also on how easily the model can interpret it. Therefore, we propose the following conjecture: \textbf{The easier the representation is to parse, the better the performance of the neural model.}

We draw on formal language theory to measure the parsing difficulty of the representations. The above representations can be viewed as a formal language specified by a grammar. The grammars corresponding to the above example are shown in the right column of Figure~\ref{figure:intro_example}, and we provide a detailed analysis in Section~\ref{sec: representation_analysis}. The classes of language grammar can be used to evaluate parsing difficulty. For example, a language in an LL(1) class is easier to parse than a language in an LL(2) or LR(1) class, as the latter can only be parsed by an LL(2) or LR(1) algorithm, while the former can also be parsed by an LL(1) algorithm.

This paper makes the following contributions based on the conjecture and the measurement. 

\textbf{Our first contribution} is a validation of our conjecture on a small programming language.
We take the Python DSL for MathQA~\cite{amini2019mathqa, dtruong46me_mathqa_python} as subject, and design four representations for the DSL, which are in LL(1), in LL(2) and LR(1) (but not LL(1)), in LR(1) (but not LL(2)), and not context-free, in the order of increasing parsing difficulty. We further experimented on the MathQA dataset, and the result confirms our conjecture: parsing difficulty strongly correlates with model performance.


\textbf{Our second contribution} is \techname, an approach that automatically constructs an LL(1) representation for any context-free grammar.
Based on the findings in the above experiment, a grammar in LL(1) should lead to high performance. We further design an algorithm, \techname, that (1) converts any context-free grammar into LL(1), and (2) derives two translators between programs in the original grammar and those in the new grammar. This way, we automatically create an LL(1) representation for any context-free language. Converting to LL(1) may increase the representation length, reducing the model efficiency. 
\techname introduces a novel hierarchical conflict elimination method, which allows a flexible trade-off between syntactic simplicity and representation length by controlling the number of layers in conflict elimination.



\textbf{Our third contribution} is the comprehensive experimental evaluation of \techname on Python and Java, which demonstrates its effectiveness.
We have applied \techname to Python and Java, and compared our approach with the representations of the plain text, grammar-rule-based representation~\cite{liang2025grammar, sun2020treegen, grammart5,l2s,abstract}, and SimPy~\cite{simpy}. The result shows that our approach outperforms all baselines. Since a full LL(1) representation leads to longer sequences, we also create another representation that resolves conflict in only one layer. This 1-layer LL(1) representation achieves almost the same performance as the full LL(1) representations, and the sequence lengths are close to the original representation, striking a balance between effectiveness and efficiency. 


\textbf{Our fourth contribution} is a systematic classification and analysis of existing code representations within our experimental framework, offering further support for our conjecture.
We analyze the grammar classes of the existing representations. The result is still consistent with our conjecture: the parsing difficulties of the representations strongly correlate with the model performances, further validating our conjecture.

The rest of the paper is organized as follows. 
Section~\ref{sec:grammar_hierarchy} provides background on grammar classes and parsing difficulty.
Section~\ref{sec: small_language} conducts the validation on a small programming language. Section~\ref{sec:approach} introduces our approach, \techname. 
Section~\ref{sec:setup} describes the experimental setup. Section~\ref{sec:results} presents the experimental results and corresponding analysis. Section~\ref{sec:related_work} reviews related work, while Section~\ref{sec:threats} and Section~\ref{sec:conclusion} discuss threats to validity and draw a conclusion separately. 

\section{Grammar Classes and Parsing Difficulty}~\label{sec:grammar_hierarchy}

\begin{figure} 
    \centering
    \includegraphics[width=0.5\linewidth]{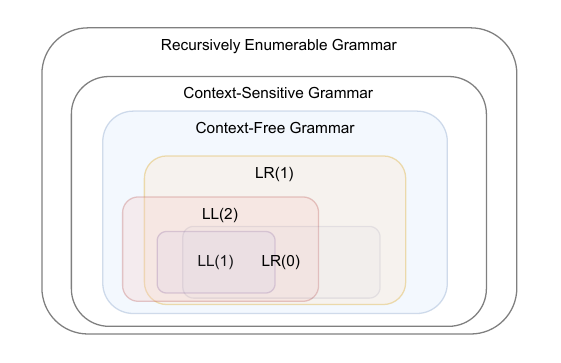}
    \caption{The hierarchical structure of grammars}
    \label{figure:grammarclass}
\end{figure}
Existing studies have recognized multiple grammar classes, each requiring a different type of parsing algorithm. 
The Chomsky hierarchy~\cite{chomsky1956three} classifies formal grammars into four classes: regular grammar, context-free grammar~(CFG), context-sensitive grammar~(CSG), and recursively enumerable grammar. As illustrated in Figure~\ref{figure:grammarclass}, these four classes of grammars are arranged in a strict inclusion hierarchy~(with regular grammars being a subset of CFGs, though not explicitly depicted). Most modern programming languages belong to the class of context-free languages. Within the context-free languages, LL(k) and LR(k) are two important families of grammar classes, representing languages parsable by an LL(k) parser and an LR(k) parser, respectively. Each LL(i)/LR(i) is a subclass of LL(i+1)/LR(i+1), and each LL(i) is a subclass of LR(i).

Larger grammar classes require more complex parsing algorithms. As a result, we can use the grammar classes to measure the parsing difficulty. Since a grammar can belong to multiple grammar classes, we use the set of grammar classes containing a grammar to represent its parsing difficulty. A grammar $A$ is considered \emph{more difficult to parse} than grammar $B$ if the set of grammar classes containing $B$ is a superset of the set of grammar classes containing $A$. For example, a grammar in LL(1) is easier to parse than a grammar in LR(1), but not in LL(1), as the former belongs to both LL(1) and LR(1), but the latter belongs only to LR(1).


\section{Validation on a Small Language}\label{sec: small_language}



In this section, we conduct an experiment to validate our core conjecture that the easier the representation language is to parse, the better the performance of the neural model. We design a controlled experiment using a domain-specific language (DSL) for mathematical expressions, varying parsing difficulty.

\begin{figure}[t]
\centering
        \includegraphics[width=\textwidth]{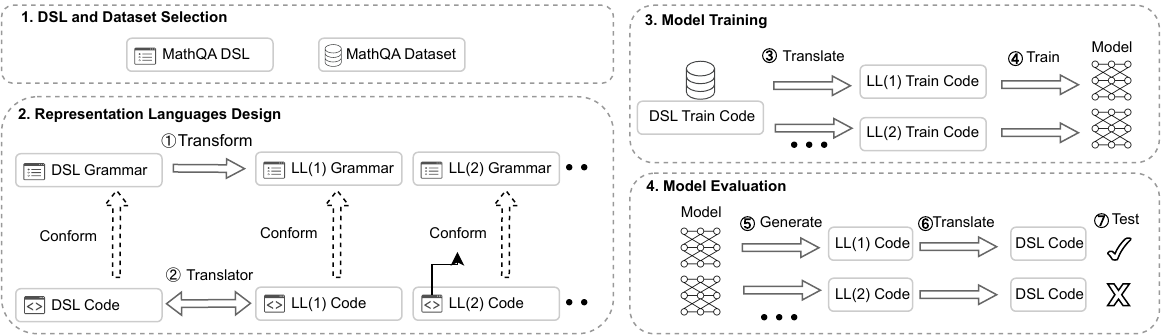}
        \caption{Overview of validation on a DSL.}
        \label{figure: dsl_workflow}
\end{figure}

Figure~\ref{figure: dsl_workflow} illustrates the experimental workflow, which consists of four phases: DSL and dataset selection, different representation language design, model training, and model evaluation. In the selection phase, we decide the experiment language and dataset. In the language design phase, we begin by transforming the DSL grammar into grammars belonging to different classes, thereby creating representation languages with distinct parsing difficulties. We then build translators that map the original DSL code to its counterparts in these new representation languages. In the model training and evaluation phase, we adopt the workflow of prior code representation approaches~\cite{grammart5, liang2025grammar, sbt, simpy}. The original DSL train code is translated into new representation languages for training, while during evaluation, the model's outputs are translated back into the original DSL code for evaluation.

\subsection{DSL and Dataset Selection}
We evaluate our approach on mathematical expression generation using the MathQA dataset~\cite{amini2019mathqa}. MathQA is a mathematical reasoning dataset that requires models to generate domain-specific language (DSL) code for solving word problems. The DSL consists of mathematical operations and variable assignments expressed in an expression-level syntax.

The dataset contains 19,209 training samples and 1,883 test samples. Each sample includes a natural language problem description followed by the corresponding DSL solution code. Figure~\ref{figure:mathqa}.a shows a typical problem where lines 1-4 describe the mathematical scenario and lines 5-6 provide the numerical values. The target DSL code (Figure~\ref{figure:mathqa}.b) follows a structured pattern: variable definition from the problem context (lines 2-4), mathematical computations (lines 5-7), and result assignment to the answer variable. We use the open-source Python version of this dataset~\cite{dtruong46me_mathqa_python} without any additional training data.

\subsection{Different Representation Languages Design}
\begin{figure}[t]
    \centering
    \begin{minipage}{0.6\textwidth}
        \centering
        \includegraphics[width=\linewidth]{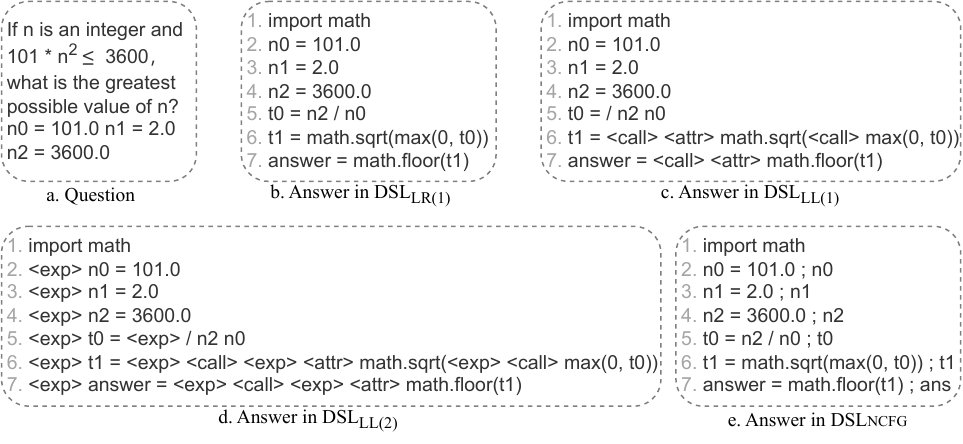}
        \caption{An example of the MathQA dataset~(<x> is new terminal)}
        \label{figure:mathqa}
    \end{minipage}
    \hfill
    \begin{minipage}{0.39\textwidth}
        \vspace{5pt}
        \centering
        \includegraphics[width=0.9\linewidth]{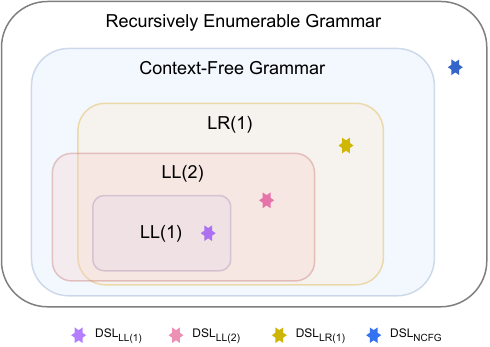}
        \vspace{5.5pt}
        \caption{Grammar hierarchy of DSLs}
        \label{figure:dsl}
    \end{minipage}
\end{figure}

The Python DSL for MathQA contains a subset of Python grammar focused on mathematical expressions, excluding complex constructs like loops, conditionals, functions, and classes. To test our conjecture about parsing difficulty and model performance, we design four representation languages with increasing parsing complexity: $\text{DSL}_{\text{LL(1)}}$ (LL(1) grammar, easiest to parse), $\text{DSL}_{\text{LL(2)}}$ (LL(2) and LR(1) but not LL(1)), $\text{DSL}_{\text{LR(1)}}$ (LR(1) but not LL(2)), and $\text{DSL}_{\text{NCFG}}$ (non-context-free grammar, hardest to parse). 
Figure~\ref{figure:dsl} shows their positions within the grammar hierarchy. Our design principle is to make minimal modifications to the original grammar while ensuring each variant belongs to its target grammar class. We modify only terminal symbols in the grammar rules, maintaining a bijective mapping between the original code and each representation language. The following sections detail the design of these languages and the translation mechanisms between them.

\subsubsection{$\text{DSL}_{\text{LR(1)}}$}
We extract the Python grammar rules that are used in the Python code in the MathQA dataset, and discard all unused grammar rules. The extracted grammar serves as the grammar of the first representation language and belongs to the LR(1) class, i.e., $\text{DSL}_{\text{LR(1)}}$. This grammar also serves as the starting point for designing other representations. 
Below we present some examples of the grammar rules.

\begin{flushleft}
\footnotesize
\centering
\(
\begin{array}{l}
\text{primary\_expression} \rightarrow \text{call} \mid \text{binary\_operator} \mid \text{attribute} \\
\text{attribute} \rightarrow  \text{primary\_expression} \, . \, \text{identifier} \\
\text{call} \rightarrow \text{primary\_expression}\ \text{argument\_list} \\
\text{binary\_operator} \rightarrow \text{primary\_expression}\ * \ \text{primary\_expression}
\end{array}
\)
\end{flushleft}
\subsubsection{$\text{DSL}_{\text{LL(1)}}$}
We transform $\text{DSL}_{\text{LR(1)}}$ into LL(1) class. This transformation differs from traditional compiler theory approaches. In compiler design, grammar transformation aims to preserve the exact language (set of strings) while changing only the grammar structure. Our transformation, however, changes the surface representation while maintaining a bijective correspondence between original and transformed programs.

The core requirement for LL(1) grammars is that for each non-terminal, all production rules must be distinguishable by their first terminal symbol. This enables LL(1) parsers to make deterministic parsing decisions using only one lookahead token. We achieve this by repositioning unique terminals to the beginning of rules and introducing new distinguishing terminals when necessary.



In the $\text{DSL}_{\text{LR(1)}}$ example, the last three rules all begin with ``primary\_expression'', which violates the principle of LL(1). After our transformation, the new grammar rules are as follows:

\begin{flushleft}
\footnotesize
\centering
\(
\begin{array}{l}
\text{primary\_expression} \rightarrow \text{call} \mid \text{binary\_operator} \mid \text{attribute} \\
\text{attribute} \rightarrow \langle \text{attribute} \rangle\ \text{primary\_expression} \, . \, \text{identifier} \\
\text{call} \rightarrow \langle \text{call} \rangle\ \text{primary\_expression}\ \text{argument\_list} \\
\text{binary\_operator} \rightarrow *\ \text{primary\_expression}\ \text{primary\_expression}
\end{array}
\)
\end{flushleft}

By introducing new terminals (such as $\langle \text{call}\rangle$ and $\langle \text{attribute}\rangle$ ) and adjusting the positions of existing terminals~(*), the new grammar rules can now be distinguished by their first symbol, i.e., satisfy the requirements of the LL(1) parser. In addition, Figure~\ref{figure:mathqa}.c presents the code example in $\text{DSL}_{\text{LL(1)}}$.

\subsubsection{$\text{DSL}_{\text{LL(2)}}$}
LL(2) parsers extend LL(1) parsers by using two lookahead symbols instead of one, enabling them to handle more complex language structures that cannot be resolved with single-token lookahead.

To create an LL(2) grammar from $\text{DSL}_{\text{LL(1)}}$, we introduce a systematic ambiguity that requires exactly two tokens to resolve. We add a common prefix symbol to related grammar rules, forcing the parser to examine both the first and second symbols to make parsing decisions. This creates a "categorize-then-select" structure where the first symbol narrows down to a rule category, and the second symbol determines the specific rule.


The grammar rule examples in $\text{DSL}_{\text{LL(1)}}$ are converted as follows:

\begin{flushleft}
\footnotesize
\centering
\(
\begin{array}{l}
\text{primary\_expression} \rightarrow \text{call} \mid \text{binary\_operator} \mid \text{attribute} \\
\text{attribute} \rightarrow \langle \text{exp} \rangle\ \langle \text{attribute} \rangle\ \text{primary\_expression} \, . \, \text{identifier} \\
\text{call} \rightarrow \langle \text{exp} \rangle\ \langle \text{call} \rangle\ \text{primary\_expression}\ \text{argument\_list} \\
\text{binary\_operator} \rightarrow \langle \text{exp} \rangle\ *\ \text{primary\_expression}\ \text{primary\_expression}
\end{array}
\)
\end{flushleft}
In this example, we attach a new terminal symbol ``$\langle \text{exp} \rangle$'' to all primary expressions. Figure~\ref{figure:mathqa}.d presents the code example in $\text{DSL}_{\text{LL(2)}}$.

\subsubsection{$\text{DSL}_{\text{NCFG}}$}
The above DSLs are all context-free languages; we also construct a non-context-free language $\text{DSL}_{\text{NCFG}}$ to represent the most complex parsing scenario. We introduce a context-sensitive constraint that requires the assigned variable to be explicitly repeated after each assignment expression. This creates a dependency where the structure depends on the specific content of variables, which exceeds the expressive power of context-free grammars and can be formally proven using the pumping lemma for context-free languages. Due to the complexity of the corresponding grammar rules, we omit the formal grammar specification here. Figure~\ref{figure:mathqa}.e presents the code example in $\text{DSL}_{\text{NCFG}}$.
\subsubsection{Code Translation}
We built a translator to translate code between the original DSL and the new DSLs. The translator performs the conversion in three steps:
(1) parse the original DSL code into its syntax tree;
(2) map this syntax tree to the corresponding syntax tree of the new DSLs;
(3) linearize the new syntax tree back into code.
The reverse translation can be carried out symmetrically.

\subsection{Model Training}

\subsubsection{Model Selection}
We choose the Qwen2.5~1.5B model~\cite{yang2024qwen2.5} for our experiments, based on two considerations. On the one hand, the model achieves a very low initial score on the MathQA dataset~(below 10\%), which helps ensure fairness by avoiding excessive prior knowledge. On the other hand, as a general-purpose model, it demonstrates a good understanding of simple math problems and can quickly adapt to this task.

\subsubsection{Training procedure}
We first translate the MathQA training set into the four representations. Then we fine-tune the Qwen2.5~1.5B model using the LLaMA-Factory framework~\cite{llamafactory2023} on these representations separately. Training is conducted on an H20 server with 6 GPUs, using a global batch size of 192 and a learning rate of 2e-5. Each model is trained for 2000 steps to ensure convergence, with checkpoints saved every 50 steps.

\subsection{Model Evaluation}

Evaluation is conducted on the MathQA test set using the pass@1 metric~\cite{chen2021evaluating}. For each question, the model generates a single program in the corresponding representation, which is then translated into Python and executed to obtain an answer. The pass@1 score is defined as the fraction of questions for which the generated program produces the correct answer, i.e.,
\[
\text{pass@1} = \frac{\text{Number of questions with correct answers}}{\text{Total number of questions}}
\]

To determine the final performance, we evaluate all checkpoints and record the pass@1 score of the top 5. We also report the mean and standard deviation of the pass@1 scores.

\begin{table}[htbp]
  \centering
  \footnotesize
  \caption{Experimental results for four DSLs}
  \begin{tabular}{lccc}
  \toprule
    \textbf{Language} & \textbf{pass@1~(Top 5 Checkpoints)(\%)} & \textbf{Mean(\%)} & \textbf{Std(\%)} \\
  \midrule
    $\text{DSL}_{\text{LL(1)}}$ &[81.89, 82.00, 82.00, 82.05, 82.05]& 82.00 & 0.07 \\
    $\text{DSL}_{\text{LL(2)}}$ &[81.68, 81.73, 81.73, 81.78, 81.78]& 81.74 & 0.04 \\
    $\text{DSL}_{\text{LR(1)}}$ &[80.99, 81.04, 81.15, 81.20, 81.31]& 81.14 & 0.13 \\
    $\text{DSL}_{\text{NCFG}}$ & [80.35, 80.35, 80.40, 80.46, 80.51] & 80.41 & 0.07 \\
  \bottomrule
  \end{tabular}
  \label{tab:mathqaresult}
\end{table}

Table~\ref{tab:mathqaresult} presents the experimental results. Different representations have a tangible impact on the model's code generation performance, exceeding random variance. Among the different representations, $\text{DSL}_{\text{LL(1)}}$ achieves the highest score (82.00), followed by $\text{DSL}_{\text{LL(2)}}$ (81.74), then $\text{DSL}_{\text{LR(1)}}$ (81.14), and finally $\text{DSL}_{\text{NCFG}}$ (80.41). Pairwise one-sided Welch’s \textit{t}-tests with Holm correction confirmed the order, showing that LL(1) outperforms LL(2) ($p = 1.66 \times 10^{-4}$), LL(2) outperforms LR(1) ($p = 1.00 \times 10^{-4}$), and LR(1) outperforms NCFG ($p = 3.69 \times 10^{-5}$). This order is consistent with the relative parsing difficulty of these representations. The results support our conjecture: the easier the representation is to parse, the better the performance of the neural model. Moreover, the results suggest that transforming the representation towards an LL(1) language may improve the model’s code generation performance.

\section{Proposed Approach}\label{sec:approach}

Based on the findings in the previous experiment, which demonstrate that easier-to-parse representations improve neural model performance, we observe that LL(1) representations may potentially lead to optimal results. Building on this insight, we propose an approach, \techname, that automatically constructs an LL(1) representation for a given context-free grammar.

\subsection{Overview}

\techname consists of two components. (1) An automatic LL(1) grammar transformation approach, which enables the conversion of an input grammar into its corresponding LL(1) version. (2) A program translator, which maps programs between the original and the transformed grammar bidirectionally. To utilize \techname within neural network models, the transformation algorithm is first employed to derive the corresponding LL(1) grammar. Subsequently, programs are translated into the new representation for model training. After the model has been trained to generate code in the new representation, the outputs can be mapped back to the original grammar using the translation procedure for downstream applications.

\subsection{Automatic LL(1) Grammar Transformation}

The core challenge in creating an LL(1) representation lies in handling LL(1) conflicts that prevent deterministic parsing. An LL(1) grammar, by definition, contains no LL(1) conflicts. LL(1) conflict arises when two production rules of the same non-terminal generate strings beginning with the same terminal. Therefore, transforming a grammar into the LL(1) class requires \textbf{detecting} and \textbf{resolving} all potential LL(1) conflicts.



A naive approach to conflict detection would compute the set of possible leading terminals for each rule through successive expansions, then check whether two rules of the same non-terminal share common terminals. However, this direct strategy proves infeasible for two reasons: (1) left recursion may cause the expansion process to never terminate, and (2) even without left recursion, the resulting abundant and intricate conflicts make subsequent resolution extremely difficult. 

To overcome these limitations, \techname adopts a hierarchical algorithm that incrementally increases the depth of expansion, identifying and resolving left recursion and potential conflicts progressively during the expansion process rather than attempting full expansion upfront. For conflict resolution, \techname employs strategies similar to those demonstrated in Section~\ref{sec: small_language}, adding new symbols or reordering existing symbols. To streamline the process, \techname first resolves conflicts only through symbol introduction, and subsequently removes redundant symbols when reordering is taken into account.

\begin{figure*}
    \centering
    \includegraphics[width=\textwidth]{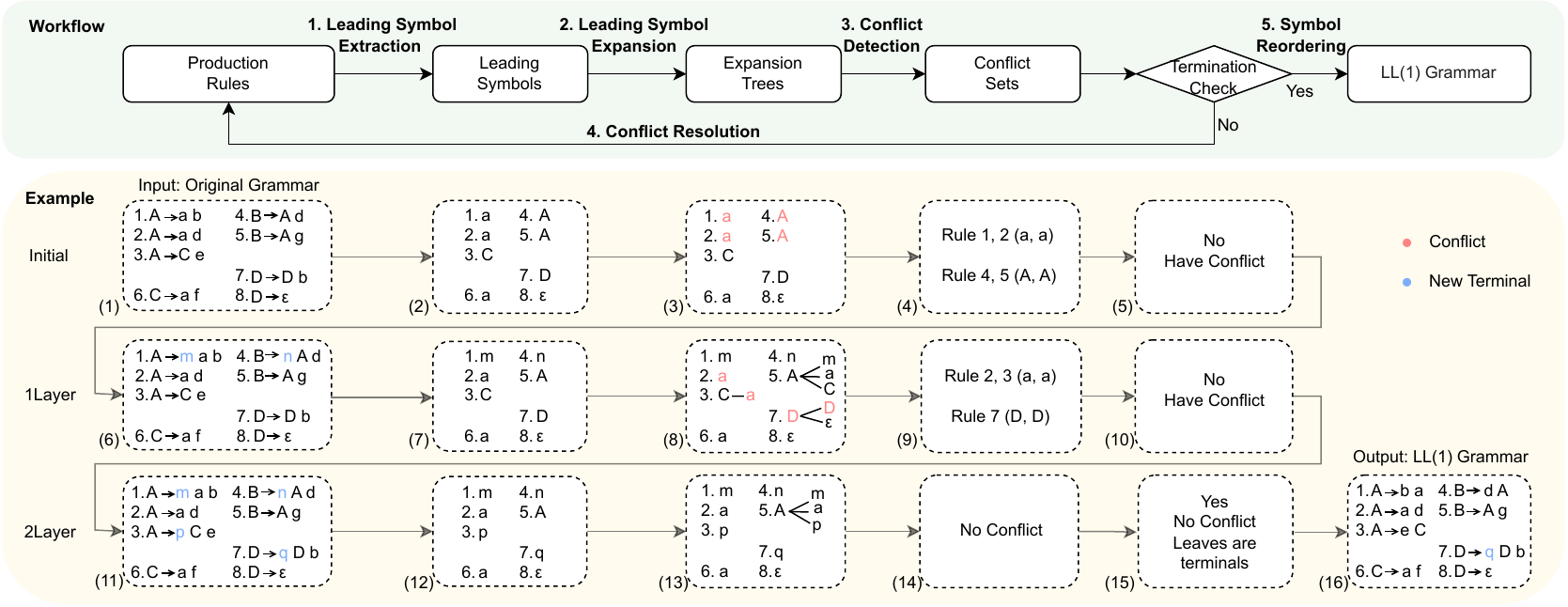}
    \caption{The workflow and an example of grammar transformation. Columns in the example are aligned with the corresponding elements of the workflow.}
    \label{figure: approach}
    
\end{figure*}

The workflow of grammar transformation is displayed in the upper part of the Figure~\ref{figure: approach}. 
The approach iteratively detects and resolves conflicts through a loop. In the i-th iteration, the procedure includes: (1) leading symbol extraction, (2) leading symbol expansion for i times, (3) conflict detection, and (4) conflict resolving.
Once all conflicts are resolved, the loop terminates, and the grammar is further simplified by (5) reordering symbols.The following sections will provide a detailed introduction of this procedure. After that, we will present an example, and prove the properties of the proposed approach.

\subsubsection{Leading Symbols Extraction and Expansion} \label{sec: 4.2.1}

\techname employs an iterative approach to detect and resolve parsing conflicts by progressively expanding grammar rules to deeper levels. In each iteration, the method increases the expansion depth and performs conflict detection at that level, enabling systematic resolution of conflicts across different expansion levels.

The iterative process works as follows: in iteration $i$, \techname expands each production rule to depth $i$ and checks for conflicts at this level. If conflicts are detected, they are resolved by introducing new terminals or restructuring rules. Crucially, after each conflict resolution step, the expansion process restarts from the beginning because the grammar modifications may propagate changes to earlier expansion layers. However, this restart is efficient since previously resolved layers remain conflict-free—only the newly expanded layer at depth $i$ requires conflict analysis.

The expansion procedure begins by extracting the leading symbol of each production rule, then collecting all leading symbols from a non-terminal's rules into a unified set. In iteration $i$, each rule's leading symbol undergoes $i$ levels of expansion, where non-terminals are recursively replaced by all symbols in their leading symbol sets, creating multiple expansion paths that form a tree structure. Conflicts are detected within the expansion trees generated from the production rules of the same non-terminal.

In particular, we treat $\varepsilon$ as an ordinary terminal rather than as the empty string. This is because $\varepsilon$-productions can still cause internal conflicts, e.g., $A \rightarrow dBc$, $B \rightarrow \varepsilon \mid c$, where $dc$ is ambiguous. In the final grammar, all $\varepsilon$ in $\varepsilon$-productions are uniformly replaced by the same terminal. Concretely, every rule of the form $X \rightarrow \varepsilon$ is rewritten as $X \rightarrow specialterminal$.

\subsubsection{Conflict Detection}

LL(1) conflicts occur when a parser cannot uniquely determine which production rule to apply for a non-terminal based on a single lookahead token. This happens when multiple production rules of the same non-terminal can generate strings that begin with the same terminal symbol, making the parsing decision ambiguous.

\techname detects such conflicts by extending beyond traditional terminal-based analysis to include leading non-terminals in the detection scope. This broader approach is based on the principle that if two productions can expand to the same leading non-terminal, they can produce the same leading terminal, indicating a parsing conflict. This strategy allows \techname to identify and resolve most conflicts in early iterations when the grammar structure is simpler, streamlining the overall transformation process. This approach motivates our use of ``leading symbols'' to encompass both terminals and non-terminals, contrasting with the conventional first sets in compiler theory that focus exclusively on terminals.

\techname identifies two primary types of conflicts during expansion tree analysis. The first type occurs when production rules of same non-terminal share identical leading symbols, manifesting as same nodes within the expansion trees of a non-terminal. When such overlapping paths are detected, \techname records the corresponding rules as conflicting since they cannot be distinguished by a single lookahead token.

The second type involves left recursion, which presents a fundamental parsing challenge because the next terminal symbol alone cannot determine the appropriate number of rule expansions. Left recursion also causes infinite expansion loops, making early detection critical for termination. In the expansion tree representation, left recursion appears as symbol repetition along any root-to-leaf path. When \techname encounters such cyclic patterns, it immediately flags the involved rules as conflicting and prepares them for resolution in the subsequent step.

\subsubsection{Conflict Resolution} \label{sec: conflict_resolution}

\techname resolves conflicts through strategic terminal insertion, employing different strategies based on the conflict type. For conflicts arising from shared leading symbols, \techname introduces a new terminal at the beginning of one of the conflicting rules, creating a unique distinguishing prefix that enables unambiguous parsing decisions. For left recursion conflicts, the approach differs: \techname adds a new terminal to the beginning of the left-recursive rule, effectively making the recursion depth explicit in the token sequence and eliminating the parsing ambiguity.

Since multiple resolution strategies exist for any given set of conflicts, \techname aims to minimize the number of new terminals introduced. When multiple conflict collections share common rules, strategic selection can resolve multiple conflicts simultaneously. For instance, consider three rules where conflicts exist between rules (1,2) and (2,3). Modifying rule 2 alone resolves both conflicts, requiring only one new terminal instead of two separate modifications.

\techname formalizes this optimization as a minimum hitting set problem: given a collection of conflicts, find the smallest subset of rules such that modifying these rules (by prepending new terminals) eliminates all conflicts. Each conflict must contain at least one modified rule to be resolved. This combinatorial optimization problem can be solved using established approximation algorithms, such as Fredman–Khachiyan algorithm~\cite{minimal3}, though the small scale of typical grammar conflicts often permits exact solutions. The resulting terminal assignments ensure that all conflicts are resolved while minimizing the grammatical complexity introduced by the transformation process.



\subsubsection{Symbol Reordering}\label{sec: reordering}

The iterative conflict resolution process terminates when no conflicts are detected in the current iteration and all leaves in the expansion trees are terminals. This condition guarantees that the resulting grammar is LL(1) by definition, as all parsing decisions can be made deterministically with single-token lookahead.

While the transformation successfully eliminates conflicts, it often introduces numerous new terminals that may be unnecessary. \techname applies symbol reordering as a post-processing optimization to reduce the number of added terminals by leveraging existing terminals that can serve the same disambiguation purpose.

\techname applies symbol reordering based on two scenarios: First, if a terminal appears uniquely in a single production rule throughout the entire grammar, that rule can be rewritten by moving the unique terminal to the front, potentially eliminating the need for newly introduced disambiguation terminals. For example, in rule $A \rightarrow dBc$ where $d$ is newly added and $c$ is unique to this rule, the rewriting $A \rightarrow cB$ achieves the same disambiguation effect. Second, when a terminal appears in multiple production rules, \techname randomly selects at most one of these rules and moves the shared terminal to the front position. 

This optimization maintains the LL(1) property since the reordered rule still begins with a unique terminal that deterministically identifies the production choice, while reducing the complexity of the transformed representation.

\subsubsection{An Example}
The lower part of Figure~\ref{figure: approach} illustrates an example of grammar transformation. 
The production rules of the initial grammar are shown in the upper-left corner, consisting of eight rules over four non-terminals. 

In iteration~0, where the leading symbols are expanded zero times, two conflicts arise: the same terminal ``a'' in rules~1 and~2, and the same non-terminal ``A'' in rules~4 and~5. 
To resolve them, two new terminals ``m'' and ``n'' are added at the beginning of rules~1 and~4, respectively. 

In iteration~1, with one expansion of leading symbols, two further conflicts appear: the same terminal ``a'' in rules~2 and~3, and left recursion in rule~7. 
These are resolved by adding terminals ``p'' and ``q'' at the beginning of rules~3 and~7. 

In iteration~2, the leading symbols are expanded twice. 
All leaves are terminals, and no conflicts remain. 
The loop terminates, yielding an LL(1) grammar with four newly introduced terminals (``m'', ``n'', ``p'', and ``q''). 
After reordering the terminals ``b'', ``e'', and ``d'', only ``q'' remains as a new terminal. 
The final LL(1) grammar is presented in the bottom-right corner.

\subsubsection{Properties}
We now show that our algorithms have the following two properties. 

\noindent \textbf{Property 1.} The output grammar is guaranteed to be LL(1).

By the definition of an LL(1) grammar, the first terminal symbol is sufficient to determine the unique production rule of any non-terminal. 
Our algorithm ensures this property as follows: for each non-terminal, the sets of leading terminals of different production rules are guaranteed to be non-empty (ensured by Section~\ref{sec: 4.2.1}) and pairwise disjoint (ensured by Sections~\ref{sec: conflict_resolution} and~\ref{sec: reordering}). 
Therefore, for any given lookahead terminal, there exists exactly one applicable production rule that satisfies the definition of LL(1).

\noindent \textbf{Property 2.} The transformation preserves a one-to-one correspondence between the syntax trees of programs conforming to the original grammars and those conforming to the target grammar.

Each syntax tree uniquely corresponds to a sequence of production rules in its leftmost derivation. Thus, it suffices to show that every valid sequence of production rules in the source grammar~(i.e., one that yields a syntax tree via leftmost derivation) can be mapped to a unique valid sequence in the target grammar, and vice versa. This property is ensured by the transformation process, because (1) every production rule in the original grammar has a unique counterpart after transformation, which allows any source sequence to be mapped to a target sequence, and (2) the transformation modifies only terminal symbols while keeping all non-terminals unchanged, which guarantees that the mapped sequence remains valid.

\subsection{Translations Between Programs in Two Grammars}\label{sec: translate}

Based on the proven one-to-one correspondence between the original and transformed syntax trees, we can construct a bidirectional translator that converts programs between the two representations while preserving their semantic meaning.

The translation process operates through syntax tree manipulation. Given a source program, we first parse it using the appropriate grammar to construct its syntax tree. We then transform this tree by mapping each production rule application to its corresponding rule in the target grammar. Since our transformation only affects terminal symbols while preserving the non-terminal structure, this mapping process updates the terminal nodes according to the established correspondence between production rules. The resulting syntax tree conforms to the target grammar and can be serialized back into a program string. 

\subsection{Partial Usage of \techname}
Under the hierarchical conflict-elimination procedure of \techname, when conflicts are eliminated to k layers, the resulting grammar~(denoted as k-layer) can still be used as a code representation. This usage balances between reducing LL(1) conflicts and the introduction of new terminals. It is worth noting that this usage may introduce potential ambiguities after terminal reordering in Section~\ref{sec: reordering}, which should be checked with the grammar parse tools such as Tree-sitter~\cite{zhang2020treesitter}.



\section{Experimental Setup}
\label{sec:setup}

In this section, we present the overall experimental setup. We begin with the research questions, followed by the training corpus and base models. We then describe key implementation and training details, and conclude with the benchmarks and evaluation metrics used in our experiments.


\subsection{Research Questions}
To comprehensively evaluate the effectiveness of the \techname, we ask the following research questions.

\paragraph{RQ1} \textit{Does \techname improve model performance on code generation tasks compared to other code representation approaches?}

We apply our approach to Python, the most common programming language in existing code generation benchmarks, to answer this question.
We apply \techname to transform Python into the LL(1) class (denoted as Python\textsubscript{LL(1)}), and also construct a partial variant by resolving only one layer of conflicts (denoted as Python\textsubscript{1-layer}). We compare them against several widely used baselines under the same experimental settings, including plain text~(denoted as Python), grammar-rule-based representations~(denoted as Python\textsubscript{Grammar}), and SimPy~(denoted as SimPy)~\cite{simpy}. This allows us to evaluate the effectiveness of our newly constructed representations in a realistic and widely adopted setting.


\paragraph{RQ2} \textit{Can \techname generalize to other programming languages?}

Building on the findings from the previous research question, we further apply \techname to Java by constructing a partial LL(1) representation of Java~(denoted as Java\textsubscript{1-layer}). Using the same training and evaluation setup, we compare the performance of the new representation against the plain Java string on the HumanEval-X benchmark. This allows us to examine the cross-language applicability of our method and validate its effectiveness beyond a single programming language.

\paragraph{RQ3} \textit{Does the parsing difficulty of a representation correlate with model performance?}

This research question is intended to validate our core conjecture on a broader scale. To answer this question, we classify the existing code representations based on their underlying grammar classes. We then examine whether representations easier to parse~(e.g., LL(1)) consistently lead to better performance, with a statistical significance analysis.


\begin{figure}[t]
    \centering
    \includegraphics[width=\textwidth]{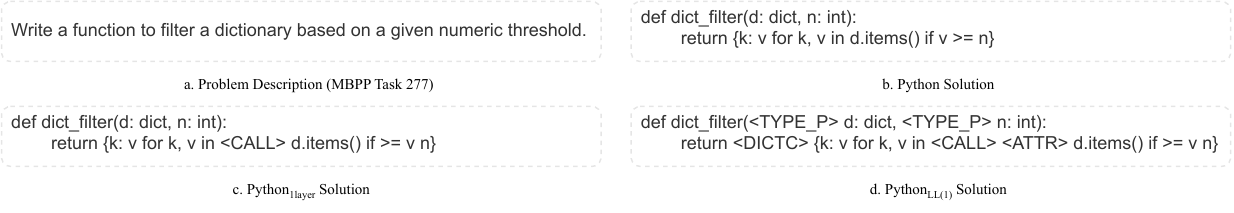}
    \caption{An example of the program in Python\textsubscript{1-layer)} and Python\textsubscript{LL(1)}~(<x> is new terminal)}
    \label{figure: python}
\end{figure}
\subsection{Implementation Details}
The experiment workflow follows a process similar to that in Section~\ref{sec: small_language}. We first converted the Python programs in the training set into different representations and trained separate models on each dataset. The models were then evaluated by generating code on the benchmarks, with the outputs translated back into Python for assessment. Translators for Python\textsubscript{LL(1)} and Python\textsubscript{1-layer} were implemented according to Section~\ref{sec: translate}. Figure~\ref{figure: python} illustrates an example of programs in Python\textsubscript{LL(1)} and Python\textsubscript{1-layer}. The translators for Python\textsubscript{Grammar} and SimPy relied on the tools provided in the original repositories. By contrast, the plain text representation required no translation; however, we applied uniform code formatting, since the translated program is consistently formatted. 

\subsection{Training Corpus and Base Models}
To enable the models to enhance these representations, we conduct instruction tuning on top of pretrained language models with a large amount of corpus. The corpus is constructed from publicly available datasets, including code-contests-instruct~\cite{beespoke2024codecontests}, Opencoder-sft-data~\cite{huang2024opencoder, opencoder2024sftstage1}, Code-290k-ShareGPT-Vicuna-Clean~\cite{banksy2024code290k}, and CodeFeedback-Filtered-Instruction~\cite{map2024codefeedback}. After removing duplicates, the dataset contains approximately 4 million high-quality instruction samples.

To ensure the generality of the experiment and mitigate the influence of model-specific variance, we select three widely adopted base models for tuning and evaluation: StarCoder~1B~\cite{li2023starcoder} as a representative of earlier open-weight code models, DeepSeek-Coder~1.3B~\cite{guo2024deepseek} as a strong code-focused model, and Qwen2.5~1.5B~\cite{yang2024qwen2.5} as a strong general-purpose foundation model. Since mastering new representations requires training on a massive corpus, the chosen models are already the largest scale that can be supported by our available resources.


\subsection{Training Details}
We conduct training on an 8-GPU H20 server using the LLaMA-Factory framework. The training is performed with a learning rate of 5e-5 and a cosine learning rate scheduler. We use a global batch size of 288 and train for 5 epochs to ensure model convergence. A checkpoint is saved every 200 steps. All experiments are carried out under the same training configuration. Due to the large size of the dataset, training one model takes approximately 5 to 6 days.

\subsection{Benchmarks}
We evaluate these models using the widely adopted HumanEval~\cite{chen2021evaluating} and MBPP~\cite{austin2021program} benchmarks, along with their enhanced versions, EvalPlus~\cite{liu2023evalplus}. HumanEval consists of 164 problems that require completing functions based on signatures and natural language docstrings, while MBPP contains 378 tasks where models are asked to generate functions based on problem descriptions and test cases. The output of models is in different representations, and we translated them into Python for testing.

\subsection{Evaluation Metrics}
We adopt pass@1 as our evaluation metric, which measures the proportion of problems successfully solved. Each model is evaluated across all its saved checkpoints. We obtain four scores corresponding to HumanEval~(+) and MBPP~(+). And then we select the top five checkpoints with the highest average score and report their mean and standard deviation as the final results. Compared to the common practice of reporting only the best checkpoint, this approach reduces the influence of randomness.

\section{Results}
\label{sec:results}

\begin{table}[t]
  \centering
  \scriptsize
  \setlength{\tabcolsep}{4pt}
  \caption{Performance of various code representations and models on HumanEval~(+) and MBPP~(+). For each model, we selected the top 5 checkpoints based on average performance, and reported the final results using their mean scores. The table shows the pass@1 scores (± standard deviation)}
  \begin{tabular}{llccccc}
    \toprule
    \textbf{Model} & \textbf{Representation} & \textbf{HumanEval} & \textbf{HumanEval+} & \textbf{MBPP} & \textbf{MBPP+} & \textbf{Avg}\\
    \midrule
    \multirow{5}{*}{StarCoder~1B}
      & Python  & 64.4 (±0.7) & 60.5 (±0.8) & 66.7 (±0.4) & 57.4 (±0.7) & 62.2 (±0.1) \\
      & SimPy   & 66.5 (±0.7) & 62.9 (±0.8) & 66.4 (±0.6) & 57.2 (±0.7) & 63.3 (±0.1) \\
      & Python\textsubscript{Grammar} & 68.7 (±1.1) & 64.6 (±0.9) & \textbf{69.6} (±1.1) & 58.5 (±1.0) & 65.4 (±0.7) \\
      & Python\textsubscript{1-layer}  & 70.1 (±0.7) & 64.8 (±1.0) & 69.3 (±0.9) & \textbf{59.1} (±0.4) & 65.8 (±0.1) \\
      & Python\textsubscript{LL(1)}   & \textbf{72.0} (±1.1) & \textbf{67.2} (±1.3) & 68.6 (±1.1) & 58.0 (±1.3) & \textbf{66.4} (±0.1) \\
    \midrule
    \multirow{5}{*}{DeepSeek\mbox{-}Coder~1.3B}
      & Python  & 66.4 (±0.5) & 62.4 (±0.9) & 72.0 (±0.7) & 60.0 (±0.6) & 65.2 (±0.1) \\
      & SimPy   & 65.2 (±1.8) & 61.0 (±0.9) & 71.5 (±0.7) & 60.2 (±1.3) & 64.5 (±0.5) \\
      & Python\textsubscript{Grammar} & 69.3 (±0.9) & 65.1 (±0.9) & \textbf{73.5} (±0.6) & \textbf{61.7} (±0.8) & 67.4 (±0.5) \\
      & Python\textsubscript{1-layer}  & 71.7 (±0.7) & 66.9 (±0.5) & 73.2 (±0.8) & 61.4 (±0.7) & \textbf{68.3} (±0.2) \\
      & Python\textsubscript{LL(1)}   & \textbf{72.3} (±1.0) & \textbf{67.8} (±1.0) & 72.6 (±0.4) & 60.3 (±0.8) & 68.3 (±0.5) \\
    \midrule
    \multirow{5}{*}{Qwen2.5~1.5B}
      & Python  & 64.9 (±0.5) & 59.6 (±0.9) & 70.3 (±0.8) & 59.0 (±0.4) & 63.4 (±0.5) \\
      & SimPy   & 67.2 (±1.6) & 62.8 (±0.7) & 70.3 (±1.0) & 59.1 (±1.1) & 64.9 (±0.4) \\
      & Python\textsubscript{Grammar} & 67.9 (±1.1) & 61.7 (±1.9) & 72.0 (±1.1) & 61.0 (±0.9) & 65.7 (±0.4) \\
      & Python\textsubscript{1-layer}  & \textbf{70.4} (±1.3) & 64.1 (±1.3) & \textbf{72.9} (±0.9) & \textbf{61.7} (±0.3) & \textbf{67.3} (±0.5) \\
      & Python\textsubscript{LL(1)}   & 69.6 (±1.8) & \textbf{64.5} (±0.7) & 72.9 (±0.7) & 61.6 (±0.5) & 67.2 (±0.4) \\
    \bottomrule
  \end{tabular}
  \label{tab:pythonresult}
\end{table}

\subsection{RQ1: Effectiveness of \techname}

Table~\ref{tab:pythonresult} presents the results of the Python experiments, covering five representations evaluated across three models and four benchmarks. The reported values are the mean and standard deviation of the top five checkpoints.

Across all three models, both Python\textsubscript{1-layer} and Python\textsubscript{LL(1)} consistently outperform the baselines (Python, SimPy, and Python\textsubscript{Grammar}) on average. For example, on StarCoder~1B, the average pass@1 improves from 62.2~(Python) to 65.8~(Python\textsubscript{1-layer}) and 66.4~(Python\textsubscript{LL(1)}). On DeepSeek-Coder~1.3B, both Python\textsubscript{1-layer} and Python\textsubscript{LL(1)} achieve the highest average score of 68.3, while on Qwen2.5~1.5B, Python\textsubscript{1-layer} achieves the best overall result (67.3), slightly outperforming Python\textsubscript{LL(1)} (67.2). Besides, Python\textsubscript{Grammar} ranks third overall, surpassing both Python and SimPy, which show comparable performance.

On individual benchmarks, Python\textsubscript{1-layer} and Python\textsubscript{LL(1)} are consistently ranked at or near the top. For example, on MBPP and MBPP+, both representations are strong, with Python\textsubscript{1-layer} slightly ahead in most cases. On HumanEval and HumanEval+, Python\textsubscript{LL(1)} often shows marginal gains, especially in StarCoder and DeepSeek. These results indicate that resolving even one layer of grammar conflict yields significant improvements, while complete conflict elimination only brings minor additional gains.

Table~\ref{tab:token_number} presents the average number of training tokens required by each representation. Both Python\textsubscript{1-layer} and Python\textsubscript{LL(1)} representations are notably more compact than the \texttt{Grammar} representation, which also outperforms Python. In contrast, Python\textsubscript{1-layer} increases token length by only 4–5\% across all models, striking a much better balance between structural clarity and sequence efficiency.
Although Python\textsubscript{1-layer} is slightly longer than the token-optimized SimPy representation~(e.g., 177 vs. 156 tokens on Qwen2.5~1.5B), it delivers significantly better performance, highlighting that preserving essential syntactic structure is more beneficial than aggressive compression. Given its superior accuracy and minimal overhead, Python\textsubscript{1-layer} is especially well-suited for large-scale training settings where both performance and efficiency matter.

Overall, both Python\textsubscript{1-layer} and Python\textsubscript{LL(1)} representations achieve better performance than existing representations. Among them, Python\textsubscript{1-layer} provides the best trade-off between effectiveness and efficiency.



\begin{table}[t]
\centering
\scriptsize
\caption{Average number of training tokens for the representations and their relative change compared to the original representation.}
\label{tab:token_number}
\begin{tabular}{lccc}
\toprule
 & \textbf{StarCoder~1B} & \textbf{DeepSeek-Coder~1.3B} & \textbf{Qwen2.5~1.5B}\\
 \midrule
Python & 192~(100.0\%) & 217~(100.0\%) & 169~(100.0\%)\\
SimPy & 179~(92.9\%) & 196~(90.5\%) & 156~(92.1\%)\\
Python\textsubscript{Grammar} & 360~(187.3\%) & 372~(172.0\%) & 347~(204.9\%)\\
Python\textsubscript{1-layer} & 201~(104.3\%) & 225~(104.0\%) & 177~(104.7\%)\\
Python\textsubscript{LL(1)} & 233~(120.9\%) & 263~(121.6\%) & 203~(120.0\%)\\
\bottomrule
\end{tabular}
\end{table}

\subsection{RQ2: Generalizability of \techname}
\begin{table}
\centering
\scriptsize
\caption{The scores of $\text{Java}$ and $\text{Java}_{\text{1-layer}}$ on the Java version of HumanEval-X and the average number of training tokens.}
\label{tab:java}
\begin{tabular}{lcccc}
  \toprule
  & \multicolumn{2}{c}{\textbf{DeepSeek-Coder~1.3B}} & \multicolumn{2}{c}{\textbf{Qwen2.5~1.5B}} \\
  \cmidrule(lr){2-3} \cmidrule(lr){4-5} 
 & \textbf{Score} & \textbf{Training Tokens} & \textbf{Score} & \textbf{Training Tokens} \\
 \midrule
Java & 58.2~(±0.7)& 159.2~(100.0\%)& 58.5~(±0.7)&122.9~(100.0\%) \\
Java\textsubscript{1-layer} & 60.0~(±1.0)& 159.3~(100.0\%)& 61.0~(±0.9)&120.0~(97.6\%)\\
\bottomrule
\end{tabular}
\end{table}
To assess the cross-language generalizability of our approach, we apply the 1-layer transformation, an effective representation as validated in previous results, to Java and evaluate its performance on the HumanEval-X benchmark~(denoted as \texttt{Java$_\text{1-layer}$}).

As shown in Table~\ref{tab:java}, \texttt{Java$_\text{1-layer}$} consistently outperforms the Java plain text across both models. On DeepSeek-Coder~1.3B, the pass@1 score increases from 58.2 to 60.0, improving by 3.1\%. On Qwen2.5~1.5B, the score improves from 58.5 to 61.0, marking a 4.3\% improvement. These gains are comparable to those observed on Python benchmarks, indicating that \techname is not language-specific but rather generalizable to other widely used programming languages.
Importantly, the improved performance is achieved without increasing the average number of training tokens. For DeepSeek-Coder, the token count remains effectively unchanged (159.2 vs. 159.3), while for Qwen2.5, the  \texttt{Java$_\text{1-layer}$} representation is even slightly more compact (120.0 vs. 122.9). This suggests that the performance improvement stems from structural simplification rather than increased input length.

Overall, these results demonstrate that our approach generalizes effectively to Java, preserving its performance advantages while maintaining high efficiency. This supports the broader applicability of \techname beyond Python, and highlights its potential as a universal representation strategy across languages.

\subsection{RQ3: Grammar Class vs. Performance}

 \begin{figure*}
    \centering
    \includegraphics[width=1\textwidth]{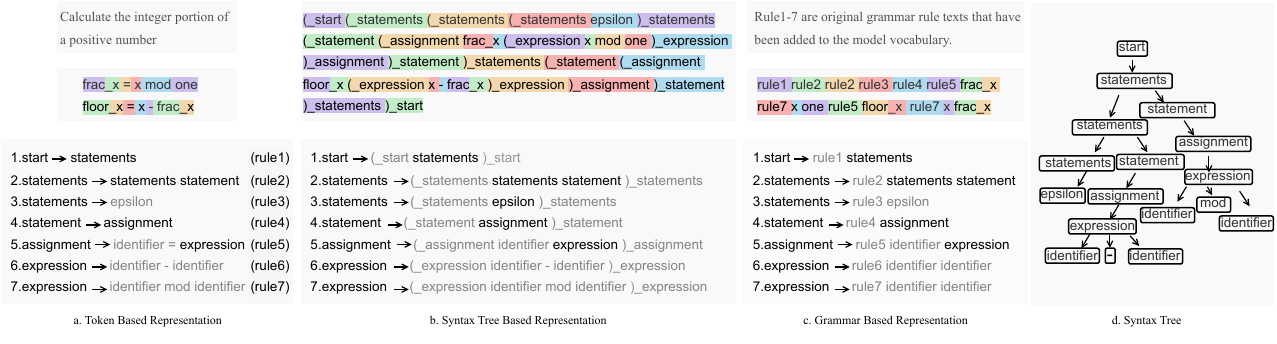}
    \caption{Grammars of Existing Representations. }
    \label{figure: analysis}
\end{figure*}

To provide a clearer and more intuitive understanding of our findings, we categorize existing code representations based on their grammar class. This analysis aims to explore whether grammar complexity correlates with model performance, offering further evidence for our core conjecture that parsing difficulty influences model understanding and generation quality.

Existing code representation approaches can be broadly categorized into four types: 
\begin{itemize}
    \item Plain Text: The program is treated as natural text, reflecting the grammar class of the standard language~(Python, Java).
    \item Syntax-Tree-Based Representation: Using a traversal of syntax tree nodes as input, these methods linearize the syntactic structure of code into a sequence. Structure-Based Traversal~(SBT) is reported to have the best performance among them~\cite{sbt}. As far as we are aware, the grammar class of SBT is still unknown.  
    \item Grammar-Rule-Based Representation: 
    Using sequences of grammar production rules as input, this line of work explicitly encodes syntactic derivations. To our knowledge, there is no analysis of the grammar class of these representations. 
    
    \item Specially Designed Programming Languages: Only with one implementation on Python named SimPy~\cite{simpy}, which has already been shown in the GLR class.
    
\end{itemize}




Due to the grammar classes of the syntax-tree-based representation and grammar-rule-based representation have not been formally specified, we first determine the grammar class corresponding to each representation. Based on this classification, we then analyze the relationship between grammar class and model performance.


\subsubsection{Analysis of Syntax-Tree-Based Representation}\label{sec: representation_analysis}

Among syntax-tree-based representations, SBT is one of the most effective and is selected as a representative for analysis.

Figure~\ref{figure: analysis}.a presents a sample code snippet and its corresponding grammar, while Figure~\ref{figure: analysis}.d shows its syntax tree. The SBT method operates by augmenting the pre-order traversal of the syntax tree: when visiting each non-terminal node, a pair of terminal symbols labeled with the node’s name is recorded at the start and end of traversal of its subtree. The final representation for this example is illustrated in the upper part of Figure~\ref{figure: analysis}.b.

We construct the grammar for the SBT-based representation by modifying the original syntax tree, inserting matching terminal nodes at the boundaries of each non-terminal's subtree.
Once the modified syntax tree is obtained, the corresponding grammar can be easily derived by extracting the parent-child relationships within it. The lower part of Figure~\ref{figure: analysis}.b presents the corresponding grammar rules.

In general, for a grammar rule of the form 
\[
  A \to B \ ,where\  A \in V ,  B \in \{V, \Sigma\}^* 
\]
The SBT method changes the grammar to
\[
A \to (_A\ B\ )_A \ ,where \ \ (_A\ , )_A\ in\ \Sigma
\]

From the derived grammar, we can analyze its parsing properties. The SBT transformation preserves the LR(1) property when the original grammar is LR(1), since the bracket structure provides sufficient lookahead information for bottom-up parsing. Besides, the resulting grammar is not LL(1) compatible. This is because for any non-terminal, all production rules begin with the same opening bracket symbol (e.g., $(_A$), creating immediate LL(1) conflicts that cannot be resolved by a single lookahead token. Furthermore, if the original grammar contains left recursion, the bracket structure introduced by SBT preserves this left recursion, which fundamentally violates LL(1) requirements for top-down parsing.

\subsubsection{Analysis of Grammar-Rule-Based Representation}
In contrast, grammar-rule-based representations are more uniform, typically generated by recording grammar rule texts during a pre-order traversal of the syntax tree. The resulting representation is illustrated at the top of Figure~\ref{figure: analysis}.c.

Similarly, we can modify the syntax tree to get the grammar for the grammar-rule-based representation. We add a new terminal node labeled with the corresponding grammar rule at the beginning of each non-terminal’s subtree. Besides, since the grammar-rule-based approach does not output terminals, we also removed the terminal symbol subnodes. The resulting grammar rules are shown in the lower part of Figure~\ref{figure: analysis}.c.

In general, for a grammar rule of the form 
\[
  A \to B \ ,where\  A \in V ,  B \in \{V, \Sigma\}^* 
\]
The grammar-rule-based method changes the grammar to
\[
A \to \langle A \to B \rangle \ B_{\text{new}} \ ,where \ \langle A \to B \rangle\  in\  \Sigma
\]
where \( B_{\text{new}} \) is derived from \( B \) by removing deterministic terminal symbols.

It is straightforward to observe that the grammar-rule-based representation falls into the LL(1) category, as each grammar rule has a unique non-terminal as its first symbol on the right-hand side.
\subsubsection{Performance Analysis}
Based on the above analysis of grammar classes, we summarise the parsing difficulty of different representation approaches in Table~\ref{tab:representation_difficulty_reason}. Furthermore, Table~\ref{tab:sigtest} reports the results of a statistical significance analysis~(one-sample t-test and the Wilcoxon signed-rank test) based on Table~\ref{tab:pythonresult}, offering a clearer view of how these approaches perform relative to the baseline method. Drawing on these two tables, we provide a comparative analysis of the approaches to validate our conjecture further.

\begin{table}[h]
\centering
\scriptsize
\caption{Comparison of representation approaches, Grammar Class,  LL(1) Conflict and parsing difficulty~(compare with Plain Text). The specific reference method is in parentheses.}
\begin{tabular}{lccc}
\toprule
\textbf{Representation Approaches}  & \textbf{Grammar Class} &\textbf{LL(1) Conflict}& \textbf{Parsing Difficulty}\\
\midrule
Plain Text (Python) & LR(1) & \textbf{Baseline}& \textbf{Baseline}\\
Syntax-Tree-Based Representation (SBT)   &  LR(1) & Increase & May increase\\
Specially Designed Language (SimPy)& LR(1) & Unchanged & Unchanged \\
Grammar-Rule-Based Representation (Python\textsubscript{Grammar}) & LL(1) & Decrease & Reduced \\
GramTrans (Python\textsubscript{1-layer})& LR(1) & Decrease& Reduced \\
GramTrans (Python\textsubscript{LL(1)})& LL(1) & Decrease & Reduced\\
\bottomrule
\end{tabular}
\label{tab:representation_difficulty_reason}
\end{table}

\begin{table}[ht]
\centering
\scriptsize
\caption{Statistical Significance of performance differences between different representation approaches and the baseline approach (Plain Text), based on the results in Table~\ref{tab:pythonresult}. Significance was tested using the one-sample t-test and the Wilcoxon signed-rank test, where $p<0.05$ indicates significance. }
\begin{tabular}{lcccc}
\toprule
\textbf{Method } &  \textbf{$t$} & \textbf{$p$ (t-test, one-tailed)} &\textbf{ $p$ (Wilcoxon)} & \textbf{Model Performance~(Compare with Plain Text)}\\
\midrule
SimPy   &  1.26 & 0.116 & 0.212 & No significant difference \\
Python\textsubscript{Grammar} &  8.64 & $1.57\times 10^{-6}$ & $2.4\times 10^{-4}$ & Significantly better\\
Python\textsubscript{1-layer}  &  7.35 & $7.23\times 10^{-6}$ & $2.4\times 10^{-4}$ & Significantly better\\
Python\textsubscript{LL(1)}   &  4.98 & $2.06\times 10^{-4}$ & $2.4\times 10^{-4}$ & Significantly better\\
\bottomrule
\end{tabular}
\label{tab:sigtest}
\end{table}



\textbf{For the syntax tree–based representation}, the representation language does not fall into a grammar class that is easier to parse; it still belongs to LR(1). Moreover, the additional symbols introduced by the transformation bring extra LL(1) conflicts, potentially increasing the overall parsing difficulty. However, we did not include this representation in our experiments because it is verbose and less effective. The existing studies~\cite{sun2023abstract} have shown that SBT performs worse on code-related tasks compared to the baseline~(plain text). This observation is in line with our conjecture.

\textbf{For the specially designed language}, the grammar class remains unchanged, and the transformation primarily focuses on reducing the number of symbols rather than lowering parsing difficulty. As a result, the parsing difficulty is almost the same as that of Python. In our experiments, SimPy does not show a statistically significant improvement over the token-sequence baseline ($p=0.116$ for the t-test and $p=0.212$ for the Wilcoxon test, both greater than $0.05$). Its performance fluctuates slightly across models but does not consistently surpass Python.

\textbf{For the grammar-rule–based representation}, we demonstrate that the representation language belongs to the LL(1) grammar class, which is easier to parse. Both previous work and our own significance analysis confirm this advantage: the representation consistently outperforms plain text ($p=1.57\times10^{-6}$ for the t-test and $p=2.4\times10^{-4}$ for the Wilcoxon test, both $<0.05$). This provides evidence in support of our conjecture.

\textbf{For the GramTrans transformations (Python\textsubscript{1-layer} and Python\textsubscript{LL(1)})}, both methods precisely eliminate LL(1) conflicts, thereby reducing parsing difficulty. The significance analysis of our experiments shows that their performance is significantly better than the token-sequence baseline (Python\textsubscript{1-layer}: $p=7.23\times10^{-6}$ for the t-test and $p=2.4\times10^{-4}$ for the Wilcoxon test; Python\textsubscript{LL(1)}: $p=2.06\times10^{-4}$ for the t-test and $p=2.4\times10^{-4}$ for the Wilcoxon test, all $<0.05$). These results are consistent with our conjecture.

Taken together, these results provide strong empirical evidence for our conjecture: the easier the representation language is to parse, the better the neural model performs.

\section{Related Work}
\label{sec:related_work}
Existing code representation approaches can be broadly classified into the following categories:

Treating programs as plain text~\cite{bpe,feng2020codebert} like natural language is straightforward. However, the syntactic structure inherent in programming languages can be overlooked, which can influence the performance of code generation.

Moreover, syntax-tree-based~\cite{alon2018general, leclair2019neural, shahbazi2021api2com, jiang2021treebert, zhou2019devign, hua2020fcca, liu2023graphsearchnet} and grammar-rule-based~\cite{sun2019grammar,sun2020treegen,liang2025grammar, liangbipartite} representations acknowledge the importance of structural information, exposing the parsed structure directly to language models. Yet, because language models operate on linear sequences, even structured information must be serialized into strings before use, and these strings also need to be parsed to recover meaning. Existing studies have not conducted a deeper analysis of this point and simply assume that directly exposing will lead to better performance.

Recently, researchers have proposed designing custom programming grammars tailored for language models. For example, SimPy~\cite{simpy} focuses on reducing token length by rewriting Python into a more compact format with fewer tokens. However, there is no research focusing on designing custom grammar to improve model performance.

In this work, we connect representation parsing difficulty to model performance, offering a better explanation of performance differences across representations, and introduce a language~(representation) design approach \techname that helps models achieve better performance.

In addition, several studies have examined the connection between large language models and the Chomsky hierarchy~\cite{deletang2022neural, dave2024investigating, yang2024masked}, treating large models as automata and analyzing their recognition abilities. In contrast, our work investigates how the parsing difficulty of representations affects code generation performance, situated in a different domain.

\vspace{4pt}
\section{Threats to Validity}
\label{sec:threats}


\textbf{Internal Validity} arises from two main factors.
The first threat comes from the limited number of DSL variants designed for the MathQA dataset, which is constrained by computational resources. To reduce bias, we selected variants with clear distinctions and strong representativeness. Furthermore, subsequent experiments on Python and Java confirm that the conclusions drawn from these DSL-based studies remain robust.
The second threat concerns the inherent randomness in model training. To mitigate this, we fix random seeds, use consistent hyperparameter settings, and ensure all experiments are conducted under identical hardware and software environments. Models, datasets, and frameworks are obtained from open repositories such as HuggingFace and GitHub. Rather than relying on single results, we report averages and standard deviations across multiple saved checkpoints.
\textbf{External Validity} potentially arises from model and benchmark selection. We evaluate three representative models: StarCoder~1B (early generation), DeepSeek-Coder~1.3B (code-specific), and Qwen2.5~1.5B (general-purpose), which also constitute the largest scale we can feasibly support, for all representation studies are conducted at this scale for consistency. To mitigate bias from benchmark selection, we adopt the most widely used datasets for code generation, HumanEval and MBPP. Together, these choices reduce threats to external validity.


\vspace{4pt}
\section{Conclusion}
\label{sec:conclusion}

In this paper, we investigated how representation structure influences code generation, guided by the conjecture that easier-to-parse representations enhance model performance. Through controlled experiments on a mathematical DSL, we demonstrated that languages belonging to simpler grammar classes achieve superior results. Building on this insight, we introduced \techname, a general transformation framework that converts any context-free language into the LL(1) class. Applying \techname to Python and Java and evaluating on HumanEval, MBPP, and HumanEval-X, we observed consistent performance improvements across multiple models while keeping token lengths manageable, surpassing existing representations. Further analysis of alternative representations reinforced our conjecture, underscoring the link between parsing difficulty and model performance.

Overall, our findings establish parsing difficulty as a key dimension in representation design and show that \techname offers a practical and general-purpose approach for advancing code generation performance.

\vspace{4pt}
\section{Data Availability}
The source code of this paper is publicly available for further research and experimentation at \url{https://anonymous.4open.science/r/GramTrans}.

\newpage
\bibliographystyle{ACM-Reference-Format}
\bibliography{refs}

\end{document}